\documentclass[aps,prd,twocolumn,groupedaddress,nofootinbib]{revtex4-1}
\usepackage{tensor}
\usepackage{amsmath}
\usepackage{amssymb}
\usepackage{mathtools}
\usepackage{graphicx}
\usepackage{wrapfig}
\usepackage{float}
\usepackage[
	colorlinks=true,
	linkcolor=blue,
	citecolor=blue,
	filecolor=magenta,
	urlcolor=blue
]{hyperref}

\begin{document}
% Title of paper
\title{Reexamining $f(R,T)$ gravity}
% Authors
\author{Sarah B. Fisher}
\author{Eric D. Carlson}
\email{ecarlson@wfu.edu}
\affiliation{Department of Physics, Wake Forest University, 1834 Wake Forest Road, Winston-Salem, North Carolina 27109, USA}
% Date
\date{\today}

\begin{abstract}
	We study $f(R,T)$ gravity, in which the curvature $R$ appearing in the gravitational Lagrangian is replaced by an arbitrary function of the curvature and the trace $T$ of the stress-energy tensor.  We focus primarily on situations where $f$ is separable, so that $f(R,T) = f_1(R) + f_2(T)$.  We argue that the term $f_2(T)$ should be included in the matter Lagrangian ${\cal L}_m$, and therefore has no physical significance.  We demonstrate explicitly how this can be done for the cases of free fields and for perfect fluids. We argue that all uses of $f_2(T)$ for cosmological modeling and all attempts to place limits on parameters describing $f_2(T)$ are misguided.  
\end{abstract}

\maketitle

\section{Introduction}
Cosmological observations have convincingly demonstrated that the expansion of the universe is accelerating \cite{Riess,Perlmutter98,Perlmutter99}.  This observation is inconsistent with Einstein's general theory of relativity for a universe containing only ordinary matter and radiation.  This suggests either the presence of novel matter with unusual properties or a breakdown of general relativity on cosmological scales \cite{Frieman}.

One modification that has gained much attention to explain this expansion is $f(R)$ gravity \cite{Buchdahl}, where the Lagrangian describing gravitational effects, normally proportional to the curvature scale $R$, is replaced by a function of that curvature.  For example, the introduction of an $R^2$ term in $f(R)$ can lead to Starobinsky inflation \cite{Starobinsky}.  The inclusion of a constant term, $f(R) = R+2\Lambda$, corresponds to the introduction of a cosmological constant, and therefore leads to the standard $\Lambda$CDM cosmology.  A further generlazation of $f(R)$ gravity was proposed by Harko {\it et al.} in \cite{Harko}, where $f(R)$ is replaced by $f(R,T)$, an arbitrary function of the scalar curvature $R$ and the trace of the stress-energy tensor $T$.  Cosmological effects of $f(R,T)$ theories have been explored by choosing several functional forms of $f$.  The separation $f(R,T) = f_1(R)+f_2(T)$ has received much attention \cite{Zaregonbadi,Velten,Carvalho,Deb,Ordines}.  In particular, for the special case $f_2(T)=-2\chi T$, limits on $\chi$ or observational predictions for non-zero $\chi$ have been applied to models of white dwarfs \cite{Carvalho}, strange stars \cite{Deb}, and Earth's atmosphere \cite{Ordines}.

As we will argue below, when we can separate these theories in the form $f(R,T) = f_1(R)+f_2(T)$, the term $f_2(T)$ should not be treated as a new contribution to the gravitational action, but instead should be incorporated into the matter Lagrangian ${\cal L}_m$. In Section II, we will introduce the formalism and discuss general principles.  In Section III, we will demonstrate how $f_2(T)$ can be incorporated into ${\cal L}_m$ for the trivial case of a free field.  In Section IV, we will show how this can be implemented for a generalized perfect fluid.  In Section V, we will summarize our conclusions and briefly discuss some of these ideas in generic $f(R,T)$ gravity.  Throughout this paper we use units where $c=1$, our metric signature is $(+---)$, and our curvature is given by $R_{\mu\nu}=R_{\mu\nu}^{MTW}$, and $R=-R_{MTW}$ where $MTW$ refers to the conventions of Misner, Thorne, and Wheeler~\cite{MTW}.

\section{$f(R,T)$ Formalism}
With our conventions, the conventional action takes the form
\begin{subequations}
\begin{eqnarray}
I&=&\int \; d^4x\sqrt{-g}{\cal L} \; ,\label{Action}\\
{\cal L} &=& {\cal L}_m - \frac{1}{2 \kappa^2} R \; , \label{StandardLagrangian}
\end{eqnarray}
\end{subequations}
where $\kappa^2 = 8\pi G$.  A cosmological constant $\Lambda$ can be included by adding $-\Lambda \kappa^{-2}$ to ${\cal L}$.  This term can be thought of as either a modification to gravity, $R \rightarrow R + 2\Lambda$, or a modification of the matter Lagrangian ${\cal L}_m \rightarrow {\cal L}_m - \Lambda \kappa^{-2}$.  The two interpretations are physically indistinguishable.

The stress-energy tensor is defined in general as\footnote{Harko {\it et al.} \cite{Harko} has this equation with the wrong sign.  This error and consequences thereof were copied by other authors.\cite{Carvalho,Velten,Deb}}
\begin{equation}\label{StressEnergy}
T^{\mu\nu}=- {2\over\sqrt{-g}}{\delta I_m\over\delta g_{\mu\nu}} \;
\end{equation}
where $I_m$ is the contribution to $I$ from ${\cal L}_m$.  Where one incorporates the contribution from the cosmological constant affects the value of the stress-energy tensor: including it in ${\cal L}_m$ adds a term $\Lambda \kappa^{-2} g_{\mu\nu}$ to $T_{\mu\nu}$.  Not surprisingly, Einstein's equations, which are derived by demanding that the full action eq.~(\ref{Action}) remain stationary under changes of the metric $\delta g_{\mu\nu}$, are identical in both cases.  Because the metric has no divergence, one can easily show that in either case $\nabla_\mu T^{\mu\nu} = 0$, so the stress-energy tensor will be conserved in both cases.

We discuss a cosmological constant here only to emphasize an important point: whether a term is inserted into the Lagrangian as matter or gravity is not physically meaningful.  In this case, because the stress-energy is conserved either way, we cannot base a decision on naturalness either. When we generalize to $f(R,T)$ gravity, we will make the case that some apparently ``new physics'' results from artificial division of the Lagrangian, and is thus non-physical.  Because the cosmological constant is not really relevant to the subsequent discussion, we will dispense with it.

The premise of $f(R,T)$ gravity, as first suggested by Harko {\it et al.}~\cite{Harko} is to replace $R$ with $f(R,T)$ in eq.~(\ref{StandardLagrangian}), so
\begin{equation}
{\cal L}={\cal L}_m-\frac{1}{2 \kappa^2} f\left(R,T\right) \; .\label{ModifiedLagrangian}
\end{equation}
We will focus on the subcase of $f(R,T)$ gravity in which we can cleanly separate the effects of curvature and matter, namely
\begin{equation}
f(R,T)=f_1(R) + f_2(T) \; .
\end{equation}
The modified Einstein's equations in this theory, derived from demanding that the action remain invariant under changes of the metric, take the form
\begin{eqnarray}
&&f_1^\prime(R) R^{\mu\nu} - \frac12 g^{\mu\nu} f_1(R) - \left(\nabla^\mu \nabla^\nu - g^{\mu\nu} \nabla^2 \right) f_1^\prime(R) \nonumber \\
&&\qquad = \kappa^2 T^{\mu\nu} + \frac12 f_2(T)g^{\mu\nu} + f_2^\prime(T) \frac{\partial T}{\partial g_{\mu\nu}} \; ,\label{ModifiedEinstein}
\end{eqnarray}
where primes denote derivatives with respect to the argument.  As has been noted in the literature, such a theory does not, in general, result in the conservation of the stress-energy tensor.  Instead we find:
\begin{equation}\label{NonConservation}
\nabla_\mu T^{\mu\nu} = - \frac{1}{\kappa^2} \nabla_\mu \left[f_2^\prime(T) \frac{\partial T}{\partial g_{\mu\nu}}\right] - \frac{1}{2 \kappa^2}  g^{\mu\nu} \nabla_\mu f_2(T)  \; .
\end{equation}

\section{Free Fields}

As a trivial example, consider a free scalar field, with matter Lagrangian
\begin{equation}
{\cal L}_m = \frac{1}{2}\left(\nabla_\mu \phi \nabla^\mu \phi -m^2 \phi^2 \right) \; .
\end{equation}
The stress-energy tensor computed from eq.~(\ref{StressEnergy}), and its trace, are then given by
\begin{subequations}
\begin{eqnarray}
T^{\mu\nu} &=& \nabla^\mu \phi \nabla^\nu \phi - \frac{1}{2}g^{\mu\nu} \left(\nabla_\alpha \phi \nabla^\alpha \phi - m^2 \phi^2 \right) \; , \label{StressEnergyFree} \\
T &=&- \nabla_\alpha \phi \nabla^\alpha \phi + 2m^2 \phi^2 \; .
\end{eqnarray}
\end{subequations}
Consider a simple linear term, where
\begin{equation}\label{Linearf2}
f_2(T) = - \frac{\kappa^2 \chi T}{4 \pi} \; .
\end{equation}
We note that, just like the cosmological constant, this contribution can be thought of as a modification of gravity or as a contribution to the matter Lagrangian.  If we view it as gravity, we find, using eq.~(\ref{NonConservation}), that stress-energy is not conserved.  We can view it as matter by defining a modified matter Lagrangian
\begin{eqnarray}
{\cal L}^\prime_m &=& {\cal L}_m + \frac{\chi T}{8\pi} \nonumber \\
&=& \frac{1}{2} \! \left( 1 - \frac{\chi}{4\pi}\right) \! \nabla_\mu \phi \nabla^\mu \phi - \frac{1}{2} \! \left(1 - \frac{\chi}{2\pi}\right)\! m^2\phi^2 \; .
\end{eqnarray}
We can then define the rescaled field and mass as
\begin{subequations}
\begin{eqnarray}
\phi^\prime &=&\phi  \sqrt{1 - \frac{\chi}{4\pi}}\; ,  \\
m^\prime &=& m \sqrt{\frac{4\pi - 2 \chi}{4 \pi - \chi}} \; ,
\end{eqnarray}
\end{subequations}
and the resultant modified matter Lagrangian reduces to
\begin{equation}\label{LScalarModified}
{\cal L}_m^\prime = \frac{1}{2}\left(\nabla_\mu \phi^\prime \nabla^\mu \phi^\prime -m^{\prime 2} \phi^{\prime 2} \right) \; ,
\end{equation}
which has the same form as the original matter Lagrangian.  Hence we see that eq.~(\ref{Linearf2}) simply rescales the field and mass.  The ``bare'' mass $m$ and field $\phi$ cannot be found in the full Lagrangian, and thus have no physical meaning.  The stress-energy tensor eq.~(\ref{StressEnergyFree}) is similarly meaningless.  A modified stress-energy tensor, derived from the modified Lagrangian eq.~(\ref{LScalarModified}), will be conserved.  The same reasoning applies to a free fermion or vector field. For more complicated functions $f_2(T)$, the resulting terms will of course not be simply a rescaling of the field, but will change the free field into an interacting field.  It will be the contention of the next section that the incorporation of $f_2(T)$ into ${\cal L}_m$ works more generally, and such incorporation should always be performed, rendering $f_2(T)$ irrelevant.

\section{Perfect Fluids}

Consider a perfect fluid that has a stress-energy tensor defined in terms of the number density of particles $n$, the entropy per particle $s$, and the fluid's local velocity vector $u^\mu$ normalized so that $u_\mu u^\mu = 1$. The particle number and entropy must be conserved, so that
\begin{subequations}
\begin{eqnarray}
0 &=& \nabla_\mu \left( n u^\mu \right) \; , \label{ConserveParticles}  \\
0 &=& \nabla_\mu \left( s n u^\mu \right) \; . \label{ConserveEntropy}
\end{eqnarray}
\end{subequations}
The stress-energy tensor is given by
\begin{equation}\label{StressEnergyPerfect}
T^{\mu\nu} = \left(\rho + p\right)u^\mu u^\nu - p g^{\mu\nu} \; ,
\end{equation}
where $\rho=\rho(n,s)$ is the energy density and  $p = p(n,s)$ is the pressure.  If stress-energy is conserved, then using the equation $u_\nu \nabla_\mu T^{\mu\nu} = 0$, one can show that the energy density and pressure are related by
\begin{equation}
n \frac{\partial}{\partial n} \rho=  \rho + p \; . \label{Pressure}
\end{equation}
When stress-energy is not conserved, we can still use eq.~(\ref{Pressure}) as a definition of $p$.  We will henceforth use this equation throughout our work without referencing it.

It is not immediately obvious how to write the matter Lagrangian for a perfect fluid.  The literature commonly assumes ${\cal L}_m = p$~\cite{Harko,Carvalho,Velten,Deb,Ordines}\footnote{Harko {\it et al.} \cite{Harko} has ${\cal L}_m = -p$ due to his error in eq.~(\ref{StressEnergy}).  This was copied by other authors.\cite{Carvalho,Velten,Deb}} without explaining where this relation comes from.  As we will demonstrate shortly, it is {\it not} true in general.  In the absence of non-standard gravity, it is derived, as done in \cite{Brown}, by taking the starting Lagrangian
\begin{equation}
 {\cal L}_m= -\rho\left(n,s\right)+J^\mu\left(\beta_{\!A}\nabla\!_\mu\alpha^A-s\nabla\!_\mu\theta-\nabla\!_\mu\phi\right) \; ,\label{FluidL}
\end{equation}
where $J^\mu = n u^\mu$,  $\alpha^A$ are a set of index functions used to label fluid flow lines, and $\beta_A$, $\theta$ and $\phi$ are Lagrange multipliers used respectively to ensure that current flows along flow lines, entropy is not transferred, and current is conserved.  The number density $n$ is now to be interpreted as an implicit function of $J^\mu$, given by
\begin{equation}\label{NumberDensity}
n=\sqrt{g_{\mu\nu}J^\mu J^\nu}\; ,
\end{equation}
and not as an independent variable.   The stress-energy tensor and its trace, computed using eq.~(\ref{StressEnergy}) will be
\begin{subequations}
\begin{eqnarray}
T^{\mu\nu} &=& (\rho + p)u^\mu u^\nu -g^{\mu\nu} {\cal L}_m \;, \label{StressEnergyFluid} \\
T &=& \rho+p - 4{\cal L}_m \;. \label{StressEnergyFluidTrace}
\end{eqnarray}
\end{subequations}

The equations of motion resulting from demanding stationarity of the Lagrangian with respect to all the fields (other than the metric) are then
\begin{subequations}
\begin{eqnarray}
0 &=& \nabla\!_\mu \left\{ \left[1 +2\kappa^{-2} f_2^\prime(T)\right] J^\mu \right\}\; , \label{ConserveN} \\
0 &=& \nabla\!_\mu\left\{s\left[1 + 2\kappa^{-2} f_2^\prime(T)\right]J^\mu\right\} \; , \label{ConserveS} \\
0 &=& \left[1 + 2\kappa^{-2} f_2^\prime(T)\right]J^\mu \nabla\!_\mu\alpha^A \; ,\\
0 & =&  -\nabla\!_\mu\left\{\beta_{\!A}\left[1 + 2\kappa^{-2} f_2^\prime(T)\right]J^\mu\right\} \; ,\\
0 &=& -\left[1 + \frac{2}{\kappa^2} f_2^\prime(T)\right]\left[\frac{\partial\rho}{\partial s}-J^{\mu}\nabla\!_\mu\theta\right] \nonumber \\
&&\qquad {} - \frac{1}{2 \kappa^2} f^\prime_2(T) {\partial \over\partial s}(\rho+p) \; ,\\
0 &=&\left[1 + \frac{2}{\kappa^2} f_2^\prime(T)\right] \left( \beta_A \nabla_\mu \alpha^A - s\nabla_\mu \theta - \nabla_\mu \phi  - \frac{\partial \rho}{\partial n} u_\mu \right)  \nonumber \\
&&\qquad {} - \frac{1}{2\kappa^2}  f^\prime_2(T) u_\mu \frac{\partial}{\partial n} \left(\rho+p\right) \; . \label{StationarityJ}
\end{eqnarray}
\end{subequations}
We can then use eq.~(\ref{StationarityJ}) to show that the Lagrangian density eq.~(\ref{FluidL}), when evaluated on shell, can be rewritten as
\begin{equation}\label{FluidLOnShell}
\overline{\cal L}_m = p + \frac{ f_2^\prime(\overline{T})}{2\kappa^2+4 f_2^\prime(\overline{T})} n \frac{\partial}{\partial n}(\rho +p) \; .
\end{equation}
where the bars on ${\cal L}_m$ and $T$ will be used to mean ``on shell" henceforward.

If $f_2=0$, then we have $\overline{\cal L}_m=p$, as is commonly assumed, and eqs.~(\ref{ConserveN}) and (\ref{ConserveS}) will conserve particles and entropy, corresponding to eqs.~(\ref{ConserveParticles}) and (\ref{ConserveEntropy}), and the stress-energy tensor eq.~(\ref{StressEnergyFluid}) will match the desired form eq.~(\ref{StressEnergyPerfect}).  But when $f_2(T)$ is anything other than a constant, this goal is not achieved, and the various terms that were included in eq.~(\ref{FluidL}) have not achieved their intended goals.  It appears that the actual conserved current should be the rescaled current $\left[1 +2\kappa^{-2} f_2^\prime(T)\right] J^\mu $.

It turns out to be more convenient to define the actual current as this taken on shell:
\begin{subequations}
\begin{eqnarray}
J^{\prime \mu} &=& \left[ 1+ 2\kappa^{-2} f_2^\prime(\overline{T})\right]J^\mu \; , \label {JPrime} \\
n^\prime &=& \left[ 1+ 2\kappa^{-2} f_2^\prime(\overline{T})\right] n \;  , \label{NPrime}
\end{eqnarray}
\end{subequations}
where the primes on $J$ and $n$ denote corrected quantities.  The change has no physical significance as all physics takes place on shell.   The on-shell trace of the stress-energy tensor $\overline{T}$ can be found as an implicit function of $n$ and $s$ by substituting eq.~(\ref{FluidLOnShell}) into eq.~(\ref{StressEnergyFluidTrace}) to yield
\begin{equation}\label{BarT}
\overline{T} = \rho - 3p - \frac{2  f_2^\prime(\overline{T})}{\kappa^2 +2 f_2^\prime(\overline{T})} n \frac{\partial}{\partial n} \left(\rho + p\right) \; .
\end{equation}
Eq.~(\ref{JPrime}) guarantees that particle number and entropy will be conserved on shell, {\it i.~e.} $\nabla_\mu J^{\prime\mu} = 0$ and $\nabla_\mu \left(s J^{\prime\mu}\right) = 0$. It is worth noting that neither the four-velocity $u^\mu$ nor the entropy per particle $s$ needs to be redefined. 

By analogy with the scalar field, we contend that the ``bare'' stress-energy tensor of eq.~(\ref{StressEnergyFluid}) is not only not conserved, it is not physically meaningful, because the separation of ${\cal L}$ into a matter term ${\cal L}_m$ and the contribution $f_2(T)$ is not physically meaningful.  Only the combination of the effects of these two quantities can be measured, and for this reason we define the physical  stress-energy tensor as
\begin{eqnarray}
T^{\prime\mu\nu}& =& T^{\mu\nu} +  \frac{1}{\kappa^2\sqrt{-g}} \frac{\partial}{\partial g_{\mu\nu}} \left[\sqrt{-g} f_2(T) \right] \nonumber \\
 &=&T^{\mu\nu} + \frac{1}{\kappa^2} f_2^\prime(T) \frac{\partial T}{\partial g_{\mu\nu}} + \frac{1}{2 \kappa^2} f_2(T) g^{\mu\nu} \; . \label{TPrimeDefinition}
\end{eqnarray}
It is easy to see from eq.~(\ref{NonConservation}) that this quantity will be conserved.

The stress-energy trace $T$, as given by eq.~(\ref{StressEnergyFluidTrace}) depends on the metric only by the implicit dependence of $\rho$ and $p$ on the number density $n = \sqrt{g_{\mu\nu}J^\mu J^\nu}$, which works out to
\begin{equation}\label{PartialT}
\frac{\partial T}{\partial g_{\mu\nu}} = \frac12 u^\mu u^\nu \left(4 + n \frac{\partial}{\partial n} \right)\left (\rho + p\right) \; .
\end{equation}
Substituting eqs.~(\ref{PartialT}) and (\ref{StressEnergyFluid}) into eq.~(\ref{TPrimeDefinition}), we find the true stress-energy tensor is
\begin{eqnarray}
T^{\prime\mu\nu} &=& \left[\rho+p + \frac{1}{2 \kappa^2} f_2^\prime(T) \left(4+n\frac{\partial}{\partial n}\right)(\rho + p) \right] u^\mu u^\nu \nonumber \\
&&\qquad {}- g^{\mu\nu} \left[{\cal L}_m  - \frac{1}{2 \kappa^2} f_2(T) \right] \; . \label{TPrimeSimplify}
\end{eqnarray}

By comparison with eq.~(\ref{StressEnergyFluid}), we see that the true energy density and pressure will be given by
\begin{subequations}
\begin{eqnarray}
\rho^\prime + p^\prime &=& \rho + p + \frac{1}{2 \kappa^2} f_2^\prime(T)\left(4+ n \frac{\partial}{\partial n}\right)(\rho + p) \; , \nonumber \\ && \label{DensityPressureTrue} \\
p^\prime &=& {\cal L}_m - \frac{1}{2 \kappa^2} f_2(T) \label{PressureTrue} \; .
\end{eqnarray}
\end{subequations}
The true density $\rho^\prime$ is the difference of these two equations, which can be simplified by using eq.~(\ref{StressEnergyFluidTrace})  to yield
\begin{eqnarray}
\rho^\prime &=& \frac14 \left(3 \rho+3p+T\right) + \frac{1}{2 \kappa^2} f_2^\prime(T) \left(4+ n \frac{\partial}{\partial n}\right)(\rho +p) \nonumber\\
&&\qquad {} + \frac{1}{2 \kappa^2} f_2(T) \; .\label{DensityTrue}
\end{eqnarray}
This equation has the disadvantage that the density is a function of all the field variables, not just $n$ and $s$.  We can correct this deficiency by replacing all the $T$'s by $\overline{T}$'s.  The formula can be further simplified by using eq.~(\ref{BarT}) to replace
\begin{equation}\label{SimplifyStep}
\frac{1}{2 \kappa^2} f_2^\prime(\overline{T}) n \frac{\partial}{\partial n} (\rho + p) = \frac14 \left[1+\frac{2}{\kappa^2} f_2^\prime(\overline{T})\right] ( \rho - 3p - \overline{T}) \, ,
\end{equation}
so we find on shell that
\begin{equation}\label{DensityPrime}
\rho^\prime = \rho + \frac{1}{2\kappa^2} \left[ f_2^\prime(\overline{T}) \left(4+n \frac{\partial}{\partial n}\right) \rho + f_2(\overline{T}) - \overline{T} f_2^\prime(\overline{T}) \right] .
\end{equation}

We have written the stress-energy tensor on shell strictly in terms of $n$ and $s$, but can we somehow incorporate $f_2(T)$ into ${\cal L}_m$, so as to eliminate the need for $f_2(T)$ entirely? Let us define a modified Lagrangian by analogy with eq.~(\ref{FluidL}), using the corrected current $J^{\prime\mu}$ and density $\rho^\prime$:
\begin{subequations}
\begin{eqnarray}
{\cal L}^\prime &=& {\cal L}_m^\prime - \frac{1}{2 \kappa^2} f_1(R) \; , \label{LPrime}\\
{\cal L}_m^\prime
 &=& -\rho^\prime\left(n^\prime,s\right)+J^{\prime\mu}\left(\beta_{\!A}\nabla\!_\mu\alpha^A-s\nabla\!_\mu\theta-\nabla\!_\mu\phi\right) \; . \nonumber \\ && \label{FluidLPrime}
\end{eqnarray}
\end{subequations}
This Lagrangian is not identical to the original Lagrangian.  But will it yield the same equations of motion?  The difference between the two Lagrangians is given by
\begin{eqnarray}
{\cal L}^\prime - {\cal L} &=& {\cal L}^\prime_m - {\cal L}_m + \frac{1}{2 \kappa^2} f_2(T) \nonumber \\
&=& \rho - \rho^\prime + \left(J^{\prime\mu} - J^\mu \right) \left( \beta_A \nabla_\mu \alpha^A - s \nabla_\mu \theta - \nabla_\mu \phi \right) \nonumber \\
&&\qquad + \frac{1}{2 \kappa^2} f_2(T)  \; . \label{LagrangianDifferenceStart}
\end{eqnarray}
This is then simplified using sequentially eqs.~(\ref{JPrime}),  (\ref{FluidL}), (\ref{StressEnergyFluidTrace}), and (\ref{DensityPrime}) to yield
\begin{equation}\label{LagrangianDifference}
{\cal L}^\prime - {\cal L} = \frac{1}{2 \kappa^2} \left[f_2(T) - f_2(\overline{T}) + \left(\overline{T} - T\right) f_2^\prime(\overline{T})\right] \; .
\end{equation}
When we apply the equations of motion, so $T = \overline{T}$, this difference vanishes.  In fact, for small variations near the stationary point, we see that
\begin{equation}\label{LagrangianDifferenceVariation}
\delta{\cal L}^\prime - \delta{\cal L} = \frac{1}{2 \kappa^2} \left\{\delta T \left[f_2^\prime(T) - f_2^\prime(\overline{T})\right] + \left(\overline{T} - T\right) \delta f_2^\prime(\overline{T})\right\} .
\end{equation}
On shell, this vanishes as well.  Hence ${\cal L}$ and ${\cal L}^\prime$ will have identical equations of motion.

Do the physical pressure $p^\prime$ and energy density $\rho^\prime$ satisfy eq.~(\ref{Pressure})?  Starting with eq.~(\ref{DensityPrime}), we see that
\begin{eqnarray}\label{PressureProof1}
n^\prime \frac{\partial\rho^\prime }{\partial n^\prime} &=& n^\prime \frac{\partial\rho}{\partial n^\prime} + \frac{1}{2 \kappa^2} f_2^\prime(\overline{T}) n^\prime \frac{\partial}{\partial n^\prime}\left(4\rho + n \frac{\partial \rho}{\partial n} \right) \nonumber \\
&&+\frac{1}{2 \kappa^2} \!\left[ n^\prime \frac{\partial}{\partial n^\prime} f_2^\prime(\overline{T})\right] \! \left(4 \rho + n \frac{\partial \rho}{\partial n} - \overline{T}\right) . \label{RhoPrime1}
\end{eqnarray}
We can now use eq.~(\ref{NPrime}) to show that
\begin{equation}\label{NVsNPrime}
n^\prime \frac{\partial}{\partial n^\prime} =\left[1 - \frac{2}{\kappa^2+2 f_2^\prime(\overline{T})} n^\prime \frac{\partial}{\partial n'} f_2^\prime(\overline{T}) \right] n \frac{\partial}{\partial n} \; .
\end{equation}
Applying this to the first two terms of eq.~(\ref{PressureProof1}) and substituting eq.~(\ref{BarT}) for $\overline{T}$ in the final term, yields, after considerable simplification,
\begin{equation}\label{PressureProof2}
n^\prime \frac{\partial\rho^\prime }{\partial n^\prime} = \rho + p + \frac{1}{2 \kappa^2} f_2^\prime(\overline{T})\left(4+ n \frac{\partial}{\partial n}\right)(\rho + p) \; .
\end{equation}
Comparison with eq.~(\ref{DensityPressureTrue}) shows that on shell we have
\begin{equation}\label{PressurePrime}
n^\prime \frac{\partial\rho^\prime }{\partial n^\prime} = \rho^\prime + p^\prime \; .
\end{equation}
Indeed, we would expect this relationship, since the true stress-energy tensor $T^{\prime\mu\nu}$ is conserved.

\section{Conclusions and Generalization}
As we have demonstrated, when $f(R,T)$ gravity can be broken into a curvature term and a stress-energy term, $f(R,T) = f_1(R) + f_2(T)$, we can incorporate $f_2(T)$ into ${\cal L}_m$ so as to eliminate the need for $f_2(T)$ entirely, and therefore $f_2(T)$ is not physically meaningful.  We demonstrated this explicitly for both a free scalar field and a generalized perfect fluid.  But sources \cite{Carvalho,Ordines} have claimed to put limits on $f_2(T)$; specifically, on the parameter $\chi$ that appears in the linear case eq.~(\ref{Linearf2}).  If it is physically meaningless, how can such limits be obtained?

These papers have made two errors.  First, they assume that ${\cal L}_m = p$, even though this formula does not generally apply.  Secondly, they identify $\rho$ and $p$ with the physical energy density and pressure.  This error is most clear in \cite{Carvalho}, which uses the equation of state for a degenerate electron gas.  The presence of a term of the form eq.~(\ref{Linearf2}) simply rescales the field and mass for a fermion, and hence the equation of state (once rescaled masses are used) for the electron gas will be unchanged.  Similarly, in \cite{Ordines} the ideal gas law is used, but this ideal gas law should be applied to the physical energy density and pressure, not the ``bare'' energy density and pressure.

Can we generalize these conclusions to generic $f(R,T)$ gravity?  Consider, for example, a perfect fluid, for which eqs.~(\ref{ConserveN}-\ref{StationarityJ}) need to be modified by replacing $f_2^\prime(T)$ with $f_T(R,T)=\frac{\partial}{\partial T}f(R,T)$.  This would suggest that the physical current, analogous to eq.~(\ref{JPrime}) should be defined as
\begin{equation}
J^{\prime\mu} = \left[1+2\kappa^{-2} f_T(R,\overline{T})\right] J^\mu \; .
\end{equation}
We note that this would result in a number density depending on the curvature, and this in turn would result in an energy density and pressure that also depend on the curvature.  Unlike $f_2(T)$, cross-terms in $f(R,T)$ will yield new physics, and limits on such terms could be placed by comparison with observations.

At the least, it seems sensible that terms in $f(R,T)$ that do not depend on curvature should be incorporated into ${\cal L}_m$, so that we could define
\begin{equation}
{\cal L}^\prime_m = {\cal L}_m - \frac{1}{2 \kappa^2} f(0,T) \; . \label{MinimalCorrection}
\end{equation}
After all, $f(0,T)$ terms represent the behavior of matter in the absence of curvature, and hence should not be considered part of the gravitational Lagrangian.  This is exactly what we did in eq.~(\ref{LScalarModified}) for a scalar field.  For a perfect fluid, this was not exactly what we did, but eq.~(\ref{LagrangianDifference}) shows that it matches what we did on-shell and to first order nearly on-shell.  This does not give us sufficient insight about how to deal with cross terms, so we do not expect that eq.~(\ref{MinimalCorrection}) would represent the correct ``physical'' Lagrangian.  General $f(R,T)$ gravity is therefore a focus of our ongoing research.
	
% \begin{acknowledgments}
%	We would like to thank 
% \end{acknowledgments}


\end{document}